\documentclass[twocolumn, prl]{revtex4}

\begin{document}
\title{A single measurement of a quantum many-body system of bosons}
\author{Zbyszek P. Karkuszewski\footnote{e-mail: zbyszek@pamir.if.uj.edu.pl}}
\affiliation{Los Alamos National Laboratory, Los Alamos, New Mexico 87544}

\begin{abstract}
Here I propose an approximate way of simulating the outcomes of a single-experiment density 
measurement that is performed on a state of $N$ bosons. The approximation is accurate if 
occupation of single-particle modes is macroscopic. 
\\ \\
LAUR 05-4019
\end{abstract}
\maketitle

\section{Introduction}
In order to simulate a single measurement of a position of a quantum particle 
described by a wave function $\phi(x)$ it is enough to randomly draw a 
position $x=\xi$ with the probability density $|\phi(x)|^2$. 
The outcome of the measurement, the particle at a point $x=\xi$, 
is different from the statistical average over many such measurements, 
given by $|\phi(x)|^2$. This kind of difference, trivial in the case of 
measurements performed on a single particle, becomes especially interesting 
in the case of quantum systems that consist of many bosons. One prominent
example is given by two colliding Bose-Einstein condensates \cite{Ketterle}, 
where every single measurement of the system reveals an interference pattern 
while no pattern is present in the average over many measurements. 

For a generic quantum state of a many-body system it is inexpensive to come
up with a prediction for an average outcome of many measurements. On the other
hand, finding out possible results of a single measurement is extremely
difficult. If the state is given by a many-body wave function
$\phi(x_1,...,x_N)$, where $x_i$ stands for a coordinate of an $i$-th
particle, one needs to draw a set of $N$ positions (one for each particle) from
the $N$-dimensional\footnote{Here and in the rest of this note we consider
one-dimensional particles} probability density $|\phi(x_1,...,x_N)|^2$. For
large number of particles, $N$, the direct sampling of the corresponding
multidimensional probability density is very difficult, if possible at all.
The direct sampling of the $N$-dimensional probability density can be replaced by
sampling of $N$ one-dimensional conditional probability densities \cite{Javanainen}. 
In practical applications, however, this clever method is suitable to handle large number of
particles only if they occupy very few modes.

In the following I propose an approximate method of simulating outcomes of
single measurements. It is designed for those quantum many-body states that
involve many macroscopically occupied modes.

A reader, novice to the subject, may establish a necessary background by contemplating 
two first pages of \cite{Javanainen} and three first pages of \cite{Rzazewski}.

Suppose that a state $|\psi\rangle$ of a system of $N$ bosons is spanned on
$M$ orthonormal modes $u_i(x)$, so that the bosonic field operator can be 
written as
$$
\hat \Psi(x)=\sum_{i=1}^M \hat a_i u_i(x),
$$
where $\hat a_i$ annihilates a boson from the mode $u_i(x)$.
The joint probability density for $N$ bosons in the state $|\psi\rangle$
is given by
\begin{equation}
p^{(N)}_{|\psi\rangle} = \frac{1}{N!}\langle \psi|\hat\Psi^\dagger(x_1)...
\hat\Psi^\dagger(x_N)\hat\Psi(x_N)...\hat\Psi(x_1)|\psi\rangle.
\label{prob}
\end{equation}
This is equivalent to the modulus squared of the $N$-body wave function.
The task is to repeatedly generate a set of $N$ numbers $\xi_1, ..., \xi_N$
according to (\ref{prob}).
As noticed by Javanainen and Yoo \cite{Javanainen}, the probability density
above
can be decomposed into a product of one-dimensional conditional probabilities
\begin{equation}
p^{(N)}_{|\psi\rangle} =
p^{(1)}_{|\psi\rangle}(x_1)p^{(2)}_{|\psi\rangle}(x_2|\xi_1)...
p^{(N)}_{|\psi\rangle}(x_N|\xi_{N-1},...,\xi_1).
\label{decomp}
\end{equation}
If one desires to generate a set of $N$ positions according to the probability
$p^{(N)}_{|\psi\rangle}$, it is enough to first pick a position $x_1=\xi_1$ with
the probability $p^{(1)}_{|\phi\rangle}(x_1)$ then a position $x_2=\xi_2$ with the 
$p^{(2)}_{|\phi\rangle}(x_2|\xi_1)$ and so on.

Each one-dimensional probability density of (\ref{decomp}) takes the following functional form 
\begin{equation}
 p^{(r)}_{|\psi\rangle}(x)=\sum_{i,j=1}^M c^{(r)}_{ij}u^*_i(x)u_j(x),
\label{spd}
\end{equation}
where the coefficients ${c^{(r)}_{ij}=c^{(r)}_{ji}}^*$ are calculated from (\ref{prob}) with
$x_1=\xi_1,...,x_{r-1}=\xi_{r-1}$. 

The important observation \cite{Javanainen,Images} is that for large $N$ the coefficients
$c^{(r)}_{ij}$ assume approximately
constant values after certain critical number $N_{crit}$ of  positions
have been drawn, i.e. $c^{(r)}_{ij}=\mbox{const}_{ij}(r)$ for  $r>N_{crit}$.
Assuming that $N_{crit}\ll N$, these constant values vary from one position measurement of $N$
particles to another. In other words, first $N_{crit}$  particles
determine the shape of the one-dimensional probability densities
$p^{(r)}_{|\psi\rangle}$.
The explanation of this "localization" of
values of the coefficients is provided in the following.

\section{The critical number of atoms.} 
A condensate is the quantum many-body state of $N$ bosons in the form $(\hat
c^\dagger)^N|0\rangle$, with $\hat c^\dagger$ denoting a creation operator of a
particle in a single-particle wave function $c(x)$.

Let us consider a superposition of two
condensates
$$
|\psi\rangle = (\hat c_1^\dagger)^N|0\rangle+(\hat c_2^\dagger)^N|0\rangle,
$$
where the operators $\hat c_1$ and $\hat c_2$ annihilate a boson in normalized single-particle
wave functions $c_1(x)$ and $c_2(x)$, respectively. These wave functions need not be orthogonal. 
An unimportant normalization factor of
$|\psi\rangle$ is omitted. Let us also define a spatial overlap of $c_1$ and $c_2$ 
\begin{equation}
o\equiv\int |c_1(x)||c_2(x)|\mbox{d}x.
\label{overlap}
\end{equation}
This quantity assumes values from the interval $[0,1]$; 
it vanishes when the wave functions are spatially separated and is equal to 
1 when $|c_1(x)|=|c_2(x)|$. In the later case, a density measurement does not
distinguish the condensates and the corresponding coefficients (\ref{spd})
would never converge to constant values. Bellow we focus on the cases
where $|c_1(x)|\ne |c_2(x)|$.

The $N$-dimensional probability density for this
state is given by
\begin{eqnarray}
p^{(N)}_{|\psi\rangle} &\propto& \prod_{i=1}^N|c_1(x_i)|^2 +\prod_{i=1}^N|c_2(x_i)|^2 
+ 2 \prod_{i=1}^N |c_1(x_i)||c_2(x_i)| \nonumber\\ 
&\times& \cos\left (\sum_{j=1}^N(\arg(c_1(x_j))-\arg(c_2(x_j)))\right ).
\label{probsep}
\end{eqnarray}
One sees that the total probability due to the last term in (\ref{probsep}) is
limited from above by $2o^N$, while the first and the second term give 2. 
Thus, if $|c_1(x)|\ne |c_2(x)|$ then $o<1$ and $o^N\to 0$ when $N$ increases,
so the last term can be neglected.
The remaining two terms define two sectors of the $N$-dimensional space where
the probability density is substantial. Again, if $|c_1(x)|\ne |c_2(x)|$ then these
sectors are spatially separated for large enough $N$. It can be easily 
seen, by extending the definition of the overlap (\ref{overlap}) to multidimensional products
of single-particle wave functions, $\int \prod_{i=1}^N|c_1(x_i)||c_2(x_i)|\mbox{d}x_i$. Such an overlap is given by $o^N$ and vanishes exponentially fast
with $N$.
Finally, if one randomly chooses a point from the $N$-dimensional space
according to the probability (\ref{probsep}), the point will belong to
the sector described by either the first term or to the sector described by the
second term in (\ref{probsep}). Thus, one samples either the condensate in $c_1(x)$
or the condensate in $c_2(x)$.

The natural requirement for this separation of the two sectors is the small value of the overlap
in $N$-dimensions. Namely, $o^N\le 0.01$, for example. The critical number
of particles saturates this inequality 
\begin{equation}
N_{crit} = \frac{\ln{(0.01)}}{\ln{(o)}},
\label{ncrit}
\end{equation}
where the number 0.01 is arbitrary (must be less than 1) and due to
logarithm in the numerator of (\ref{ncrit}), the $N_{crit}$ is not sensitive to its precise value.
The (\ref{ncrit}) provides an estimate of an average $N_{crit}$ over many repeated measurements. 
The critical number of particles for a single measurement can differ significantly 
from (\ref{ncrit}).

To better understand the meaning of $N_{crit}$ consider the case of the Gaussian wave functions
$$
|c_1(x)|^2 = \frac{1}{\sqrt{\pi}\sigma}e^{-\frac{(x+L)^2}{\sigma^2}},\quad
|c_2(x)|^2 = \frac{1}{\sqrt{\pi}\sigma}e^{-\frac{(x-L)^2}{\sigma^2}}.
$$
The overlap is a function of the parameters $L$ and $\sigma$
\begin{equation}
o=e^{-\frac{L^2}{\sigma^2}}.
\label{overlapg}
\end{equation}
From (\ref{ncrit}) and (\ref{overlapg})
$$
N_{crit} = -\ln(0.01)\frac{\sigma^2}{L^2}.
$$
Reversing the problem one can ask, which Gaussians can be distinguished 
given the total number of particles $N$ and the width $\sigma$. 
The answer: these
displaced by more than $2\sigma\sqrt{-\ln(0.01)/N}$. 
So, if the state $|\psi\rangle$ was an equal superposition of all of Gaussians,
labeled by the parameter $L$, then a single measurement of positions of $N$ particles
will converge to a subset of Gaussians centered within the distance $\sigma\sqrt{-\ln(0.01)/N}$
from some random point $L=L_0$.

Now, it is clear why the coefficients in (\ref{spd}) assume constant values
only approximately. The "localization" of their values improves with the
number of already drawn particles $r$ like $1/\sqrt{r}$.

Another important example involves two condensates that show shifted
interference-like patterns
$$
|c_1(x)|^2 = \frac{1}{2\pi}(1+\cos(x+\phi)),\quad
|c_2(x)|^2 = \frac{1}{2\pi}(1+\cos(x+\rho))
$$
for $x\in[0.2\pi]$. 
Assuming $-\pi\le\rho\le\phi\le\pi$, the overlap of $c_1$ and $c_2$ is a function of the relative
phase $\phi-\rho$
$$
o=\left(1+\frac{\phi-\rho}{\pi}\right)\cos\left(\frac{\phi-\rho}{2}\right)
+\frac{2}{\pi}\sin\left(\frac{\phi-\rho}{2}\right).
$$
For small phase shifts $\phi-\rho\ll 1$ the distinction of the two condensates
requires at least the number of particles
$$
N_{crit}\approx \frac{8\ln(0.01)}{(\phi-\rho)^2}.
$$
This example is important because $N_{crit}$ above is a critical
number of atoms for the measurement to "decide" which interference pattern is
realized in a single density measurement of two colliding condensates \cite{Ketterle,Javanainen}.

\section{The method}

The previous part of this work establishes that for $N\gg N_{crit}$ almost all the positions 
of particles  in a single simulation are randomly 
drawn from the same one-dimensional probability density. The single measurement on the state
$|\psi\rangle$ can be formally replaced by the measurement on a Bose-Einstein condensate --
the state where all the particles are described by the same single-particle wave
function. Indeed, the measurement of particle positions performed on a condensate consists on
repeated sampling of a single probability density. 

This idea leads to the following postulate:

{\it The act of position measurement on 
a many-body state $|\psi\rangle$ is equivalent to the measurement performed on a condensate
with the single-particle wave function
\begin{equation}
c(x;Q_1,...,Q_M) = \sum_{i=1}^M Q_iu_i(x),
\label{af}
\end{equation}
where $Q_i\equiv q_ie^{i\theta_i}$ are random complex parameters
satisfying the normalization condition}
\begin{equation}
\sum_{i=1}^M|Q_i|^2=1.
\label{constrain}
\end{equation}

The simulation of a single measurement of positions of $N$ particles is now reduced either to
random generation of $N$ positions, each with the probability density $|c(x; Q_1,...,Q_M)|^2$,
or to usage of the $N|c(x; Q_1,....,Q_M)|^2$ as a smoothened density of
measured particles. 

The only thing left is to provide a probability distribution of random parameters $Q_i$. 
This distribution arises from the projection of the $N$-body
state of interest on a condensate described by the wave function (\ref{af}).
The operator that annihilates a particle from the condensate mode $c(x;Q_1,...,Q_M)$ reads
\begin{equation}
\hat c = \sum_{i=1}^M Q_i^*\hat a_i.
\end{equation}
The condensate of $N$ particles is given then by
\begin{equation}
|c;Q_1,...,Q_M\rangle \equiv \frac{1}{\sqrt{N!}}(\hat c^\dagger)^N|0\rangle.
\label{condensate}
\end{equation}
Finally, the probability distribution of $Q_i$'s for the state $|\psi\rangle$
\begin{equation}
P(Q_1,...,Q_M) = {\cal N}_{|\psi\rangle} |\langle c;Q_1,...,Q_M|\psi\rangle|^2,
\label{probQ}
\end{equation}
with the constrain (\ref{constrain}). The normalization factor 
${\cal N}_{|\psi\rangle}$  has to be
computed for each state $|\psi\rangle$ separately, so that 
$\int P(Q_1,...,Q_M)\mbox{d}Q_1...\mbox{d}Q_M = 1$.

The proposed method consist of two stages: (i) random generation of
$M$ complex parameters $Q_i$ from the probability distribution (\ref{probQ}) and under
the constrain (\ref{constrain}), (ii) using this set of $Q_i$'s the wave
function (\ref{af}) is constructed and its modulus squared is taken either as a probability 
density for generating $N$ particle positions or as a smoothened density of particles in a
single measurement.

At this point three issues need to be clarified. 
First, the condensates (\ref{condensate}) form an overcomplete and nonorthogonal basis in the 
Hilbert space of all $N$-body bosonic states spanned on the modes $u_i(x)$. 
The nonorthogonality and overcompletness of the condensates are vital if
they are to describe the outcomes of measurements that are also "nonorthogonal" and "overcomplete",
in the sense that a weighted sum of two or more outcomes is also a valid
outcome. Conversely, if the basis formed by the condensates was
orthogonal then every result of measurement would be 
described by one of the basis states and the outcomes corresponding to combinations
of condensates would have been unjustly excluded.

Second, the postulate above relies on the observation that for large number of particles 
$N$ the system in a state $|\psi\rangle$ is driven to a condensate state by the position measurement.
In an actual experiment, the positions of all particles are measured at
the same instant and the initial state is destroyed due to absorption of light by particles.
It is conceivable, though, that starting from a state with macroscopically
occupied modes one can create a condensate by measuring positions of a
fraction $N_{crit}$ of particles. In the case of interacting particles, such a
condensate might be short lived since it is not an eigenstate of the
Hamiltonian of the system.

Third, the form of a possible outcome of a single density measurement (\ref{af})
resembles the wave function of a hypothetic condensate in the spontaneous
symmetry breaking approach, see \cite{Javanainen} pages 3 and 4.  
This work provides yet another justification for this approach: It gives
correct description of measurements if $N\gg N_{crit}$.

\section{Examples}

{\bf Fock state}. Lets check how the new method works for the Fock state 
$$
|\psi\rangle = |n_1,...,n_M\rangle.
$$ 

From (\ref{probQ})
\begin{equation}
P(Q_1,...,Q_M) \propto \left | Q_1^{n_1}\times ...\times Q_M^{n_M}\right |^2,
\end{equation}
where $M$ is a number of modes of the state. The probability above does not
depend on phases $\theta_i$ of $Q_i$'s, so the phases will be uniformly and independently 
drawn from the
interval $[0,2\pi)$.  Having said that, one can restrict the probability
density to the $M$-dimensional space of moduli $q_i$. 
The maximum of the probability (under the constrain 
(\ref{constrain})) is located at $q_i=\sqrt{n_i/N}$. For large number
particles per mode the probability is sharply peaked around the maximum point.
In fact, one can fix the values of moduli $q_i=\sqrt{n_i/N}$ and simply draw
the phases. Thus the expected density of particles found in a single
experiment is given by
$$
|c(x; \theta_1, ...,\theta_M)|^2 = \left |\sum_{j=1}^{M}
\sqrt{\frac{n_j}{N}}e^{i\theta_j}u_j(x)\right |^2.
$$
This example can be immediately extended to the mixtures of Fock states.

In particular, when one considers the state $|\psi\rangle = |N/2,N/2\rangle$ as in
\cite{Javanainen}, with $u_1(x)=\frac{1}{\sqrt{2\pi}}e^{ix}$ and 
$u_2(x)=\frac{1}{\sqrt{2\pi}}e^{-ix}$ one gets
the prediction for a single measurement 
$$
|c(x; \theta_1, \theta_2)|^2 = \frac{1}{2\pi}(1+\cos(x+(\theta_1-\theta_2))).
$$
Since both phases are random, their difference is also random and the results from
\cite{Javanainen} are recovered: Each density measurement will reveal
interference fringes shifted by a random phase.

%{\bf Superposition of Fock states}. The next tutorial example involves the superposition
%$$
%|\psi\rangle = |n_1, ..., n_M\rangle + e^{i\phi}|n'_1, ..., n'_M\rangle.
%$$
%Here the probability distribution takes the form
%\begin{eqnarray}
%P(Q_1,...,Q_M)&\propto & q_1^{2n_1}\times...\times q_M^{2n_M} + q_1^{2n'_1}\times
%...\times q_M^{2n'_M} \nonumber\\ 
%&+& 2q_1^{n_1+n'_1}\times ...\times q_M^{n_M+n'_M}\times  \\
%&\times &\cos((n_1-n'_1)\theta_1+... \nonumber \\
%& &...+(n_M-n'_M)\theta_M-\phi). \nonumber
%\end{eqnarray}
%One sees that now the phases $\theta_i$ are correlated 
%through the probability function. In particular, at the point of 
%maximum probability 
%$$
%(n_1-n'_1)\theta_1+...+(n_M-n'_M)\theta_M=\phi.
%$$
%In other words, all the phases are still drawn from the interval $[0,2\pi)$
%but the probability function prefers some phase configurations over the others.

{\bf Schr\"odinger cat state}.
Since the presented method has ambition to generate outcomes of a single
measurement performed on any state with macroscopically occupied modes,
it should also work for the Schr\"odinger cat state
$$
|\psi\rangle = |N,0\rangle + |0,N\rangle.
$$
If corresponding modes $u_1(x)$ and $u_2(x)$ are spatially separated, the exact calculation shows
(see the section where $N_{crit}$ is derived),
that the single measurement will reveal either
all particles in the first mode (in the state $|N,0\rangle$)
or all particles in the second mode (state $|0,N\rangle$).

The probability density for $Q_i$'s
$$
P(Q_1,Q_2) \propto q_1^{2N}+ q_2^{2N} 
+ 2q_1^Nq_2^N\cos(N(\theta_1-\theta_2)),
$$
under the constrain (\ref{constrain}), possesses two well isolated global maxima 
at $(q_1=1, q_2=0)$ and $(q_1=0, q_2=1)$ independent of the phases $\theta_i$. 
For large number of particles the probability is appreciable only in the
narrow vicinity of these global maxima. This means that one can see either
"alive" or "dead" cat in a single measurement and that superposition of "dead" and "alive"
is strongly suppressed. The approximate character of the method manifests
itself in the fact that even if the modes $u_1(x)$ and $u_2(x)$ are spatially separated,
there is nonzero probability for the superposition.

{\bf Bogoliubov vacuum}.
A more complex but very important example is the Bogoliubov vacuum state. 
For this state, there is also another method of predicting outcomes of a single density
measurement available \cite{DS}.
The Bogoliubov vacuum is especially useful to describe evolving Bose-Einstein condensates.
It can be expressed in the particle representation as
$$
|\psi\rangle = \left ( \lambda_1\hat a^\dagger_1\hat a^\dagger_1 +...
+\lambda_M\hat a^\dagger_M\hat a^\dagger_M\right )^{N/2}|0\rangle,
$$
where $\lambda_i$'s are real numbers, see \cite{Zstate,DS}.
The corresponding probability function assumes the following form
$$
P(Q_1,...,Q_M) \propto \left |\left (\sum_{i=1}^M \lambda_i Q_i^2\right
)^{N/2}\right |^2.
$$
In the case of just one depletion mode, $M=2$, $\lambda_1=1, |\lambda_2|<1$ and the
probability becomes
$$
P(\{Q\}) \propto \left ( q_1^4 + \lambda_2^2q_2^4 +
2\lambda_2q_1^2q_2^2\cos(2(\theta_1-\theta_2))\right )^{N/2}.
$$
The last expression shows a single peak at $(q_1=1, q_2=0)$ of the variance 
$(\Delta q_2)^2\propto 1/N$.

\section{Summary}
The method presented above consists on sampling of a $2M$-dimensional probability
density (\ref{probQ}). What one has to draw is a set of $M$ complex numbers $Q_i$ 
to determine an outcome of a single measurement. 
This sampling is much simpler, however, than the direct sampling of particle
positions from an $N$-dimensional probability density. First of all, for the
states with macroscopic occupation of modes $M\ll N$ and the dimensionality of
the sampling problem is strongly reduced. Second, as shown in the examples
above,
the probability distribution (\ref{probQ}) often assumes a simple
shape and either analytical calculations can be performed or the importance sampling 
numerical techniques can be applied.

The justification of the method provides also a link between sampling of
a $N$-body probability density and the corresponding spontaneous 
symmetry breaking guess for the outcome of the single density measurement.

\section{Acknowledgments}
I am grateful to Jacek Dziarmaga and Krzysztof Sacha for inspirations and
enlightening discussions. Work supported by LDRD X1F3 program.

\end{document}